\documentclass[conference]{IEEEtran}
\IEEEoverridecommandlockouts
\usepackage{cite}
\usepackage{amsmath,amssymb,amsfonts}
\usepackage{algorithmic}
\usepackage{graphicx}
\usepackage{textcomp}
\usepackage{xcolor}
\usepackage{csquotes}
\def\BibTeX{{\rm B\kern-.05em{\sc i\kern-.025em b}\kern-.08em
    T\kern-.1667em\lower.7ex\hbox{E}\kern-.125emX}}
    \makeatletter
\newcommand{\linebreakand}{%
  \end{@IEEEauthorhalign}
  \hfill\mbox{}\par
  \mbox{}\hfill\begin{@IEEEauthorhalign}
}
\makeatother

\begin{document}

\title{An Architecture for Control Plane Slicing in Beyond 5G Networks\\
}
\author{\IEEEauthorblockN{Rashmi Yadav}
\IEEEauthorblockA{\textit{Department of Electrical Engineering} \\
\textit{Indian Institute of Technology Kanpur,}\\
India \\
rashmiy@iitk.ac.in}
\and
\IEEEauthorblockN{Rashmi Kamran}
\IEEEauthorblockA{\textit{Department of Electrical Engineering,} \\
\textit{Indian Institute of Technology Bombay,}\\
India  \\
rashmi.kamran@iitb.ac.in}
\and
\IEEEauthorblockN{Pranav Jha}
\IEEEauthorblockA{\textit{Department of Electrical Engineering,} \\
\textit{Indian Institute of Technology Bombay,}\\
India\\
pranavjha@ee.iitb.ac.in}
\linebreakand 
\IEEEauthorblockN{Abhay Karandikar}
\IEEEauthorblockA{\textit{Department of Electrical Engineering,} \\
\textit{Indian Institute of Technology Bombay, India}\\
karandi@ee.iitb.ac.in\\
\IEEEauthorblockA{\textit{Director, Indian Institute Technology Kanpur, India} \\
karandi@iitk.ac.in\\
}
}

}

\maketitle
\begin{abstract}To accommodate various use cases with differing characteristics, the Fifth Generation (5G) mobile communications system intends to utilize network slicing. Network slicing enables the creation of multiple logical networks over a shared physical network infrastructure. While the problems such as resource allocation for multiple slices in mobile networks have been explored in considerable detail in the existing literature, the suitability of the existing mobile network architecture to support network slicing has not been analysed adequately. We think the existing 5G System (5GS) architecture suffers from certain limitations, such as a lack of slice isolation in its control plane. This work focuses on the future evolution of the existing 5GS architecture from a slicing perspective, especially that of its control plane, addressing some of the limitations of the existing 5GS architecture. We propose a new network architecture which enables efficient slicing in beyond 5G networks. The proposed architecture results in enhanced modularity and scalability of the control plane in sliced mobile networks. In addition, it also brings slice isolation to the control plane, which is not feasible in the existing 5G system. We also present a performance evaluation that confirms the improved performance and scalability of the proposed system viz-a-viz the existing 5G system.
\end{abstract}


\section{Introduction}
\IEEEPARstart{T} {he} emergence of the Fifth Generation (5G) mobile network enables a large variety of use cases and services. The Third Generation Partnership Project (3GPP) defined the 5G System (5GS) and categorizes its prominent use cases as Enhanced Mobile Broadband (eMBB), Ultra-Reliable Low Latency Communication (URLLC), Massive Machine Type Communication (mMTC), and Vehicle to Everything (V2X) \cite{A1}, \cite{A2}. Each of these use case categories is expected to support a different set of requirements. For example, eMBB use cases are expected to support very high data rates and low to high-speed mobility applications, while URLLC applications typically require very low latency and low to moderate data rates. Similarly, the broad characteristic of mMTC applications is to have low data rates along with very high connection density, while V2X applications require low latency support in high-speed mobility scenarios. Networks beyond the 5G era may need to support newer use cases such as holographic-type communications, tactile internet for remote operations, digital twin etc. Considering their diverse needs, each use case deserves a dedicated network infrastructure to efficiently serve the users. However, providing dedicated infrastructure to each of these use case categories may lead to an increase in Capital Expenditures and Operational Expenditures.  Hence, the concept of network slicing is adopted in 3GPP 5GS to support the use-case category-specific requirements. 

3GPP defines network slicing as “a paradigm where logical networks/partitions are created, with appropriate isolation, resources and optimized topology to serve a purpose or service category (e.g., use case) or customers (a logical system created on-demand)” \cite{RAN3}. A network slice in 3GPP 5GS spans both horizontally as well as vertically, i.e., both Radio Access Network (RAN) and Core Network (CN) and also the control and the user (data) plane functions. In addition to the requirement to support multiple network slices with isolation between them, existing 5GS needs to handle other slice-specific requirements too, as has been discussed in section 16.3 of \cite{R19}. We observe that one of these requirements, “Support for UE associating with multiple network slices simultaneously” \cite{R19}, has a significant bearing on the control plane architecture of the 3GPP 5GS. The requirement mandates that “in case a UE is associated with multiple slices simultaneously, only one signalling connection is maintained”. It implies that the control plane functions especially that terminate UE signalling, e.g., a gNB-Centralized Unit-Control Plane (gNB CU-CP) function in RAN terminating Radio Resource Control (RRC) signalling, or an Access and Mobility Management Function (AMF) in CN terminating Non-Access Stratum (NAS) signalling, may have to support more than one slice concurrently. Therefore, achieving “isolation of slices” and having “slice-specific NFs” in the control plane becomes particularly difficult. It should be noted that other 5G standards such as the one developed by O-RAN Alliance \cite{R20} also do not provide any guidance/resolution to this problem.

Network slicing in 5G networks has been an active field of research and here we provide a survey of the research work on this topic, especially those pertaining to isolation of slices. The work presented in \cite{R1} highlights the challenges of slice isolation in the user (data) plane but there is no discussion on control plane slice isolation there. Authors in \cite{R2} investigate the challenges related to RAN slice design and implementation but do not discuss slice isolation. The authors in \cite{R3} guarantee the functional and performance isolation of slices while allowing the efficient use of resources in the RAN data plane but do not discuss isolation viz-a-viz control plane. A RAN slicing architecture with multiple sets of function splits and placements, which provide isolation among slices, is proposed in \cite{R6}. However, it neither discusses control plane slice isolation nor does it discuss core network slicing. Another work \cite{R10} proposes a flexible RAN architecture with a Medium Access Control (MAC) scheduler to abstract and share physical resources among slices. In \cite{R11}, SDN enabled resource allocation framework is proposed and in \cite{R12}, a network slicing framework for end-to-end Quality of Service (QoS) with a dynamic radio resource slicing scheme is proposed. In \cite{R21}, an architecture for the cloud-network slicing concept and realization of the slice-as-a-service paradigm is presented. It is designed to consider modularity and multi-domain dynamic operation as key attributes.

As can be discerned from the literature survey presented above, most existing works focus only on the slicing aspects of the user (data) plane and to the best of our knowledge, no prior work on control plane slice isolation is found. Moreover, the architectural mechanism to enable slice isolation in the control plane of existing 5GS has neither been discussed in the standards nor in the research work.

Therefore, in order to achieve slice isolation in the mobile network control plane, we propose a new mobile network architecture in this paper. The proposed architecture improves the slice-specific design of the mobile network control plane and facilitates slice isolation therein. It is an extension of our earlier 5G-Serv architecture [16]. The 5G-Serv did not explore slicing aspects, which has been addressed in this extension. In addition, we have also done a detailed performance evaluation of the control plane in the proposed architecture in a sliced environment and compared it with the existing 5GS architecture. The performance evaluation focuses on slice-wise session establishment rate, resource utilization, scalability and modularity of the control plane. We demonstrate that the proposed architecture achieves improved control plane performance in a sliced environment as compared to the existing 5GS architecture. An additional benefit of the proposed architecture is its simplified end-user signalling viz-a-viz the 3GPP 5GS.

The rest of the paper is organized as follows: Section \ref{arch} discusses the architectural details of the proposed slice-specific control plane architecture. The procedures involved in the proposed extension are detailed in Section \ref{flow}. Section \ref{model} provides the system model. The performance evaluation is covered in Section \ref{perf}, while the conclusion is provided in Section \ref{conc}.

\section{Proposed Architecture for Control Plane Slicing }
\label{arch}
\begin{figure}[ht!]
	\centering
	\hspace{1.0cm}
	\includegraphics[width=\columnwidth]{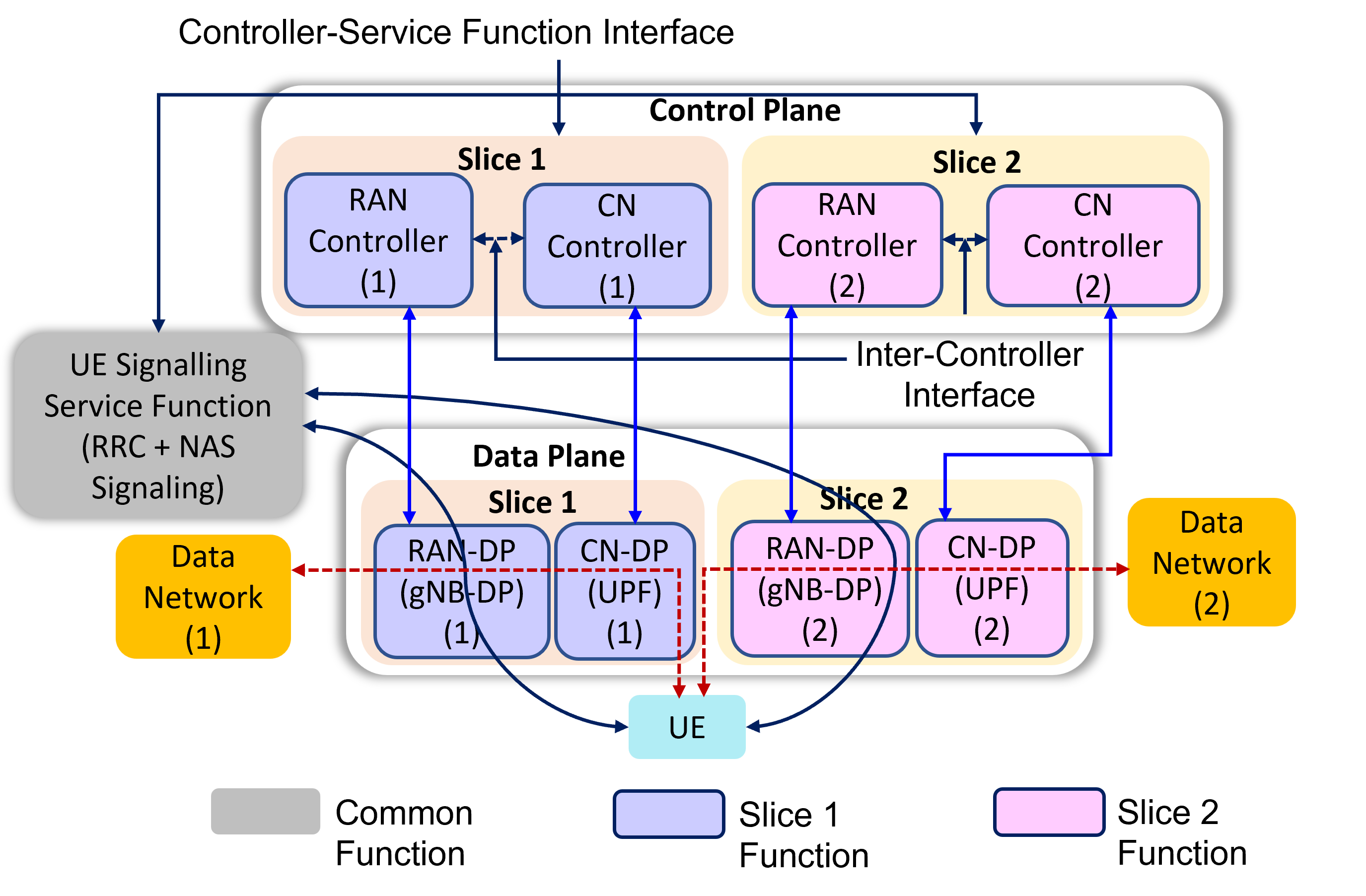}\hspace{0.2cm}
	\vspace*{-0.8cm}
	\caption{Proposed architecture for end-to-end slicing (Slice-specific functions for Slice\,1 and Slice\,2 are shown in purple and pink colors respectively, and common functions are shown in grey color).}
	\label{proposed}
	\vspace{-0.2cm}
\end{figure}

In the existing 5GS, one of the issues behind the problem of slice isolation in the control plane is the termination (placement) of UE signalling functionality within the control plane function, e.g., in gNB-CU-CP or AMF. Hence, the separation of UE signalling handling functionality from the control plane may help us in solving the problem of slice isolation in the control plane. The idea to have separate network functions for UE signalling handling, separate from the existing control plane functions, was proposed in our earlier work 5G-Serv (details available in \cite{R7}). However, it was not analysed in the context of a sliced network, which has been done in this work. 
The architecture of the proposed mobile network is divided into slice-specific control plane and data plane network functions in RAN and CN as shown in Fig. \ref{proposed}. 

In our proposal, each plane has slice-specific network functions in RAN and the core network. Besides, a few functions are slice-specific or common to a set of slices such as the UE signalling service function (comprising RRC and NAS signalling). In this work, we consider two network slices for simplicity. However, the proposed architecture can easily employ more than two slices. Various components of the proposed architecture are explained in the following sections:

\subsection {Control Plane} 
The control plane functions in the RAN and CN of the proposed architecture are named as the RAN controller and the CN controller, respectively. The RAN controller is responsible for resource allocation and data(user) plane control functionality in RAN, whereas CN controller is responsible for the same functionality in CN.
In the existing 5GS, the gNB-CU-CP, the de facto RAN control plane function, broadly contains the following functionalities\footnote{It may have some additional functionality (e.g. support for Xn interface) but those are not important for the discussion here.}: gNB-DU control, gNB-CU-UP control, RRC protocol, Radio Resource Management (RRM) and the Next Generation Application Protocol (NGAP) functionalities. We propose to change the placement of some of these functionalities to simplify the RAN control plane function for end-to-end slicing. As can be discerned, the above-mentioned functionality can broadly be divided into two classes: (i) UE-specific control/signalling functionality, e.g., UE-specific RRC protocol functionality and (ii) RAN user plane control functionality, e.g., gNB-DU/gNB-CU-UP control functionality. The UE-specific signalling functionality in gNB-CU-CP, responsible for terminating RRC protocol signalling with UEs, is moved out of the gNB-CU-CP and is transposed to a new UE signalling service function in the network. Also, UE-specific NGAP message handling, to carry NAS messages between Access and Mobility Management Function (AMF) and gNB-CU-CP in the existing 5GS, is also removed from the gNB-CU-CP. After the relocation of the UE-specific signalling handling functionality from gNB-CU-CP to UE Signaling Service Function, only the RAN user plane control functionality remains there and simplifies the overall gNB-CU-CP design. This simplified gNB-CU-CP is rechristened as the RAN controller in the proposed architecture.
\par Similarly, UE-specific signalling functionality, e.g., NAS signalling handling, UE authentication etc. is moved out of the CN control plane functions like AMF, SMF, Authentication Server Function (AUSF) and is again placed in the UE signalling service function. The remaining user plane control functionality in CN control plane is rechristened as CN controller, which is considerably simpler compared to the conventional CN control plane in the existing 5GS. The modified RAN and CN controllers can communicate through an inter-controller interface similar to the existing NGAP interface.
A key point to be noted here is that the modified RAN and CN controllers (in the proposed architecture) do not contain the UE signalling functionality and no longer terminate the UE signalling. This architectural change allows the controllers to be slice-specific, i.e., every network slice can have its own RAN and CN controllers removing the constraint for the control plane to necessarily support more than one slice in existing 5GS.

\subsection {Data (User) Plane}
\par The user plane is responsible for the transfer of data through the mobile network. There are no changes in the user plane functionality in the proposed architecture over the existing 5GS; these remain the same as the user plane functions in the existing 5GS. The gNB-Centralized Unit-User Plane (gNB-CU-UP) comprises Service Data Adaptation Protocol (SDAP), General Packet Radio Service (GPRS) Tunneling Protocol (GTP), and Packet Data Convergence Protocol (PDCP) layers, and the gNB-Distributed Unit (gNB-DU) has Radio Link Control (RLC), MAC, and Physical (PHY) layers. gNB-DU and gNB-CU-UP are altogether termed as RAN-Data Plane (RAN-DP) or gNB-Data Plane (gNB-DP). RAN-DP (gNB-DP) in RAN and the CN-DP (UPF) in CN maybe slice specific, i.e., each logical network (slice) has its RAN-DP and CN-DP (UPF) functions. 
\subsection{UE Signalling Service Function}
\par UE signalling service function exchanges signalling messages such as RRC/NAS messages with UEs. This is a new function defined as part of the proposed architecture. The CN control plane functionality in existing 5GS such as NAS signalling termination as part of AMF or UE authentication functionality as part of AUSF are moved from the CN control plane functions to the UE signalling service function in the proposed architecture as is the RRC signalling handling functionality from gNB-CU-CP. These UE signalling service functions can either be slice-specific or common to a set of slices. It is possible to have more than one UE signalling service function in the network for reasons such as load balancing and distribution of functionality across them.
\subsection{Interfaces}
\begin{figure}[h!]
	\centering
 \vspace{-0.3cm}
\includegraphics[width=0.7\columnwidth]{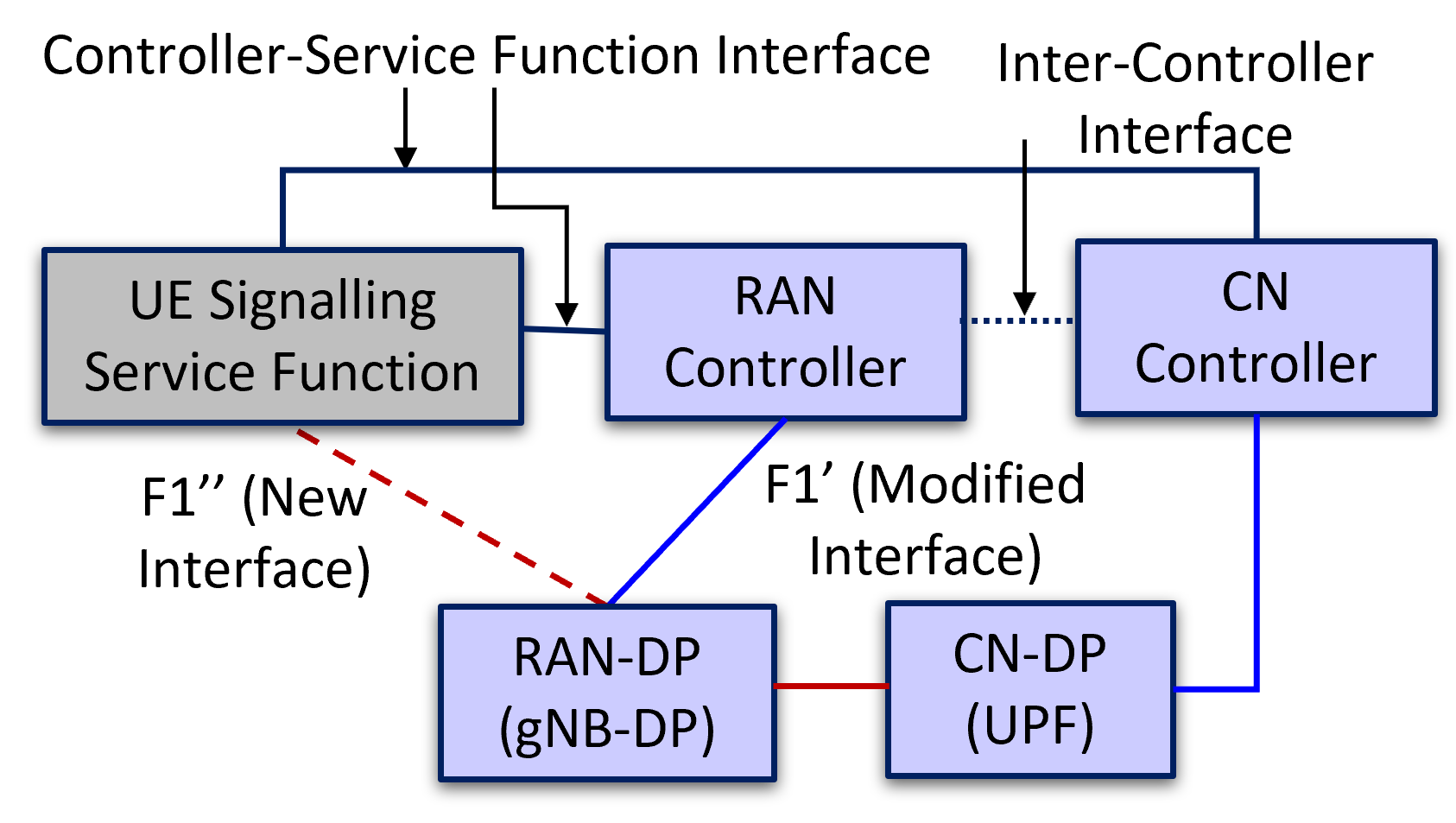}
	\vspace{-0.2cm}
 \caption{Interfaces between the RAN controller, UE signalling service function and data plane.}
	\label{interface}
	\vspace{-0.3cm}
\end{figure}

 Fig.\,\ref{interface} shows proposed/modified interfaces in the proposed architecture. gNB-CU-CP has been segregated into two different entities as RAN controller and UE signalling service function. Conventionally, F1-C is the interface between gNB-DU and gNB-CU-CP and it carries UE-specific messages. However, a modified F1-C (F1-C') interface exists between the RAN-DP and RAN Controller, which now does not carry UE-specific control messages and information elements. Besides, a new interface, F1'' is proposed between the RAN-DP and UE signalling service function, which now carries UE-specific RRC/NAS messages. An important consequence of the creation of separate UE signalling service functions, separate from the control plane functions in the proposed architecture, is that UE-specific signalling messages can be treated as another form of data passing through the user plane. Hence the proposed F1'' interface can be similar to the F1-U interface of the existing 5GS. Now, UE can exchange the signalling message with the UE signalling service function via RAN-DP over the F1'' interface. Controller-service function interface exists between the RAN controller and the UE signalling service function. Inter-controller interface is a new interface which can be based on the existing NGAP interface.
\begin{figure*}[h!]
	\centering
	\hspace{-0.3cm}
 \vspace{0.cm}
	\includegraphics[width=1.95\columnwidth]{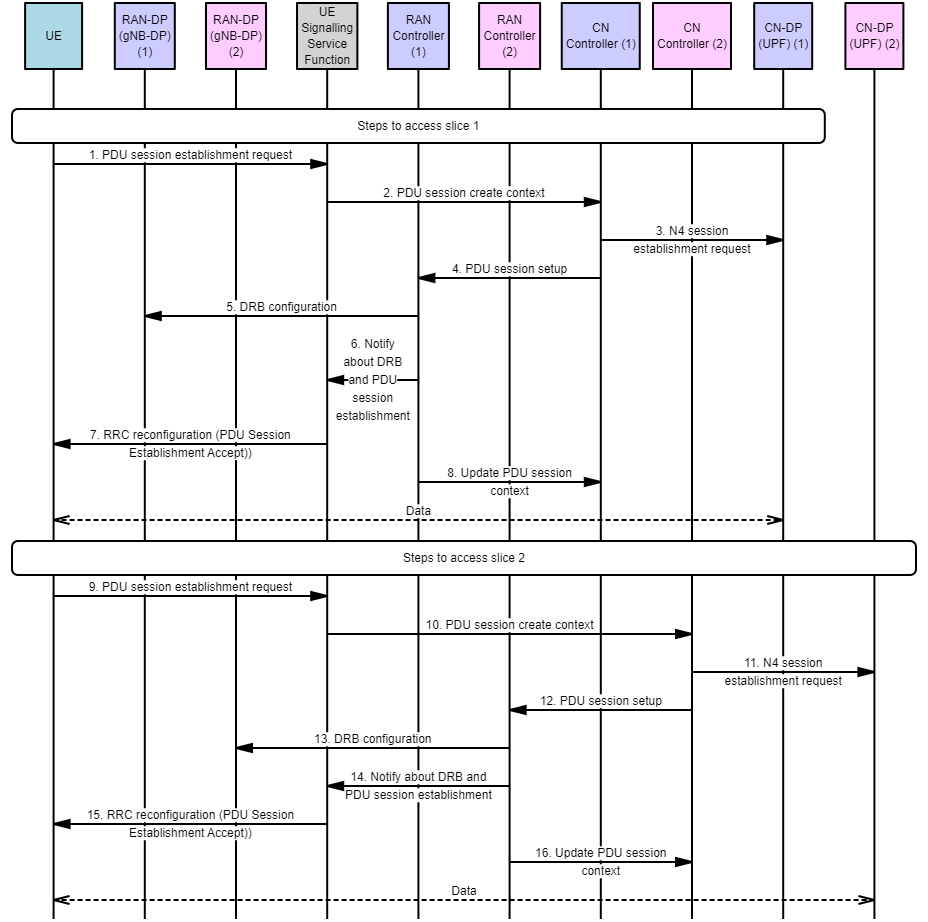}
	\vspace{-0.4cm}
	\caption{Call flow for a UE accessing two slices using the proposed architecture.}
	\label{callflow}
	\vspace{-0.4cm}
\end{figure*}
\par The proposed architecture is validated and elucidated further through an example of Protocol Data Unit (PDU) session establishment call flow in the next section.
\section{PDU Session Establishment for the Proposed Architecture} 
\label{flow}
In this section, we detail the call flow for PDU session establishment for a UE with two network slices in the proposed architecture (as shown in Fig.\,\ref{callflow}) to understand its working. Besides, we also compare it with the PDU session establishment call flow of existing 5GS in the same scenario of two slices (referred from Section 4.3.2.2 of \cite{R13}).

Firstly, UE sends the PDU session establishment request to access a network slice (say slice\,1). Based on the received request, the UE signalling service function (common to both the slices in our case) forwards the request to the respective CN controller. Further, the CN controller selects the corresponding CN-DP (UPF) for session establishment, and the N4 session is established at CN-DP (UPF). In Fig.\,\ref{callflow}, the call flow is shown for a UE connecting two network slices in which UE first connects to slice\,1 and then to slice\,2. Details of message sequences in the call flow are as follows: 
\begin{itemize}
   \item UE sends a PDU session establishment request as a NAS message to the UE signalling service function.  
    \item Based on the received request for a particular slice, the UE signalling service function selects the slice-specific controller (say CN controller-1 for the first slice) in the core network and sends a message to create a PDU session context. 
    \item Accordingly, the CN controller-1, which is specific to the slice\,1, configures CN-DP (UPF)-1. 
    \item The CN controller-1 informs the RAN controller-1 (which is also slice-specific) about the PDU session setup on the inter-controller interface. 
   \item Subsequently, the RAN controller-1 establishes a PDU session on the RAN-DP-1 by sending Data Radio Bearer (DRB) configuration message to RAN-DP-1. It also notifies about DRB and PDU session establishment to the UE signalling service function (after the PDU session is established). 
   \item  UE signalling service function sends the RRC reconfiguration message to the UE. 
    \item RAN controller-1 sends PDU session context update to CN controller-1. 
    \item PDU session is now established for UE to access slice\,1. The same message sequence is followed for UE to access the second slice, as shown in the second part of Fig.\,\ref{callflow}.
\end{itemize}

\par  On comparing the message sequences for accessing the network slices in the case of the existing 5GS \cite{R13} and the proposed system, there is a reduction in the total number of messages by employing the proposed architecture as compared to the existing 5GS. For instance, in the existing 5GS procedure, there are request/response messages sent and received between AMF and SMF for creating and updating a PDU session context. In contrast, responses from the CN controller to the UE signalling service function (for the PDU session context create) and also to the RAN controller (for the PDU session context update) are not required. N1N2 messages, which are communicated between AMF and SMF in the existing 5GS, are removed completely from the proposed call flow as the RAN controller and CN controller (SMF') can communicate directly through an inter-controller interface. Overall, the message sequence for PDU session establishment (for two slices) for the proposed scheme is simplified as compared to the existing 5GS, which validates enhanced modularity in the procedures of the proposed system.
 
\section{System Model}
 \label{model}
 In this section, we describe the system model of the proposed architecture by considering the example of PDU session establishment call flow. We used Performance Evaluation Process Algebra (PEPA) \cite{R9}, a formal high-level language for modelling distributed systems and their evaluation. The modelling of the proposed call flow (Fig.\,\ref{callflow}) is provided in Table\,\ref{tab:table1}. Various Network Functions (NFs), e.g. UE, RAN-DP-1 (DP-1), UE signalling service function (USSF), RAN controller-1 (RANC-1), CN controller-1 (CNC-1), and CN-DP (UPF)-1 for the proposed architecture are modelled as PEPA modules. The NF's states are defined with the corresponding NF name and a number ($NF_1$) (refer to Table\,\ref{tab:table1}) (e.g., $Ranc_1$ indicates the first state for the RAN controller). Further, the action types are denoted in lowercase and subscripts are added to specify the detail of the defined actions ($actiontype_{detail}$). For example, request and reply for any service, e.g. PDU session create context, can be specified as $req_{sc1}$ and $rep_{sc1}$, respectively. Each action type is associated with a specific rate value, $r$. The rates (number of actions performed per unit time) model the expected duration of a specific type of action in the PEPA component and are taken as reference from \cite{R22}, \cite{R15} and \cite{R23}.

 \begin{table}
\caption{System modelling for PDU session establishment call flow\label{tab:table1}}
\centering
\vspace{0.2cm}
\fontsize{9pt}{9pt}\selectfont
\begin{tabular}{|p{0.18\columnwidth}|p{0.7\columnwidth}|}
\hline
\textbf{PEPA Modelling of NFs} & \textbf{\centering Code Description}\\
\hline
UE & \textit{$Ue_1$ $_{=}^{def}$ ($get_{uep}$, $r_p$).($req_{se1}$, $r_{iat}$).$Ue_2$} \\
& \textit{$Ue_2$ $_{=}^{def}$ ($reconfig_1$, $r_v$).$Ue_3$} \\
& \textit{$Ue_3$ $_{=}^{def}$ ($req_{se2}$, $r_v$).($reconfig_2$, $r_v$).$Ue_1$} \\
\hline
DP-1 & \textit{$Dp_1$ $_{=}^{def}$ ($drb_1$,$r_v$).$Dp_2$} \\
NF & \textit{$Dp_2$ $_{=}^{def}$ ($get_{dpp1}$, $r_p$).($prepare$, $r_v$).$Dp_1$} \\ 
\hline
RANC-1 & \textit{$Ranc_1$ $_{=}^{def}$ ($setup_1$, $r_v$).$Ranc_2$} \\
NF & \textit{$Ranc_2$ $_{=}^{def}$ ($get_{rancp1}$, $r_p$).($drb_1$, $r_v$).$Ranc_3$} \\
& \textit{$Ranc_3$ $_{=}^{def}$ ($get_{rancp1}$, $r_p$).($notify_1$,$r_v$).$Ranc_4$} \\ 
& \textit{$Ranc_4$ $_{=}^{def}$ ($get_{rancp1}$, $r_p$).($update_1$, $r_v$).$Ranc_1$} \\

\hline
USSF & \textit{$Ussf_1$ $_{=}^{def}$ ($req_{se1}$, $r_v$).$Ussf_2$} \\
NF & \textit{$Ussf_2$ $_{=}^{def}$ ($get_{ussfp}$, $r_p$).($req_{sc1}$, $r_v$).$Ussf_3$} \\
& \textit{$Ussf_3$ $_{=}^{def}$ ($notify_1$, $r_v$).($reconfig_1$, $r_v$).$Ussf_4$} \\
& \textit{$Ussf_4$ $_{=}^{def}$ ($req_{se2}$, $r_v$).$Ussf_5$} \\
& \textit{$Ussf_5$ $_{=}^{def}$ ($get_{ussfp}$, $r_p$).($req_{sc2}$, $r_v$).$Ussf_6$} \\
& \textit{$Ussf_6$ $_{=}^{def}$ ($notify_2$, $r_v$).($reconfig_2$, $r_v$).$Ussf_1$} \\
\hline
CNC-1 & \textit{$Cnc_1$ $_{=}^{def}$ ($req_{sc1}$, $r_v$).$Cnc_2$} \\
NF & \textit{$Cnc_2$ $_{=}^{def}$ ($get_{cncp1}$, $r_p$).($req_{n4est1}$, $r_v$).$Cnc_3$} \\
& \textit{$Cnc_3$ $_{=}^{def}$ ($rep_{n4est1}$, $r_v$).$Cnc_4$} \\
& \textit{$Cnc_4$ $_{=}^{def}$ ($get_{cncp1}$, $r_p$).($setup_1$, $r_v$).$Cnc_5$} \\
& \textit{$Cnc_5$ $_{=}^{def}$ ($update_1$, $r_v$).$Cnc_1$} \\
\hline
CN-DP (UPF)-1 & \textit{$Upf_1$ $_{=}^{def}$ ($req_{n4est1}$, $r_v$).$Upf_2$} \\
NF & \textit{$Upf_2$ $_{=}^{def}$ ($get_{upfp1}$, $r_p$).$Upf_1$} \\
\hline
UE & \textit{$Uep_1$ $_{=}^{def}$ ($get_{uep}$, $r_p$).$Uep_2$} \\
Processor & \textit{$Uep_2$ $_{=}^{def}$ ($req_{se1}$, $r_v$).$Uep_1$} \\
\hline
DP-1  & \textit{$Dpp_1$ $_{=}^{def}$ ($get_{dpp1}$, $r_p$).$Dpp_2$} \\
Processor &\textit{$Dpp_2$ $_{=}^{def}$ ($prepare$, $r_v$).$Dpp_1$} \\
\hline
RANC-1 & \textit{$Rancp_1$ $_{=}^{def}$ ($get_{rancp1}$, $r_p$).$Rancp_2$} \\
Processor & \textit{$Rancp_2$ $_{=}^{def}$ ($drb_1$, $r_v$).$Rancp_1$}\\&+($notify_1$, $r_v$).$Rancp_1$+($update_1$, $r_v$).$Rancp_1$ \\
 
\hline
USSF & \textit{$Ussfp_1$ $_{=}^{def}$ ($get_{ussfp}$, $r_p$).$Ussfp_2$} \\
Processor & \textit{$Ussfp_2$ $_{=}^{def}$ ($reconfig_1$, $r_v$).$Ussfp_1$}\\
& +($reconfig_2$, $r_v$).$Ussfp_1$ \\
\hline
CNC-1 & \textit{$Cncp_1$ $_{=}^{def}$ ($get_{cncp1}$, $r_p$).$Cncp_2$} \\
Processor & \textit{$Cncp_2$ $_{=}^{def}$ ($req_{n4est1}$, $r_v$).$Cncp_1$} \\
& +($setup_1$, $r_v$).$Cncp_1$ \\
\hline
CN-DP (UPF)-1 & \textit{$Upfp_1 (_{=}^{def}) (get_{upfp1}, r_p).Upfp_2$} \\
Processor & \textit{$Upfp_2$ $_{=}^{def}$ ($req_{n4est1}$, $r_v$).$Upfp_1$} \\
\hline
System  & $Ue_1$[n] $_{\phi}^{\bowtie}$ $Dp_1$[N]) $_{\phi}^{\bowtie}$ $Dp_2$[N]) $_{S_1}^{\bowtie}$ $Ussf_1$[N] \\
Equation & $_{S_2}^{\bowtie}$ $Ranc_1$[N] $_{S_3}^{\bowtie}$ $Ranc_2$[N] $_{S_4}^{\bowtie}$ $Cnc_1$[N]  \\
& $_{S_5}^{\bowtie}$ $Cnc_2$[N] $_{S_6}^{\bowtie}$ $Upf_1$[N] $_{S_7}^{\bowtie}$ $Upf_2$[N] \\
& $_{S_8}^{\bowtie}$ $Uep_1$[n] $_{\phi}^{\bowtie}$ $Dpp_1$[$N_p$] $_{\phi}^{\bowtie}$ $Dpp_2$[$N_p$] \\
& $_{\phi}^{\bowtie}$ $Ussfp_1$[$N_p$]) $_{\phi}^{\bowtie}$ $Rancp_1$[$N_p$]) $_{\phi}^{\bowtie}$ $Rancp_2$[$N_p$]) \\ 
& $_{\phi}^{\bowtie}$ $Cncp_1$[$N_p$] $_{\phi}^{\bowtie}$ $Cncp_2$[$N_p$] \\ 
& $_{\phi}^{\bowtie}$ $Upfp_1$[$N_p$] $_{\phi}^{\bowtie}$ $Upfp_2$[$N_p$] \\
\hline
Variables & $S_1$ = $<$$req_{se1}$, $reconfig_1$, $req_{se2}$, $reconfig_2$$>$ \\ 
& $S_2$ = $<$$drb_1$, $notify_1$$>$ \\ 
& $S_3$ = $<$$drb_2$, $notify_2$$>$ \\ 
& $S_4$ = $<$$req_{sc1}$, $setup_1$, $update_1$$>$ \\ 
& $S_5$ = $<$$req_{sc2}$, $setup_2$, $update_2$$>$ \\ 
& $S_6$ = $<$$req_{n4est1}$$>$ \\
& $S_7$ = $<$$req_{n4est2}$$>$ \\ 
& $S_8$ = $<$$get_{uep}$, $get_{dpp1}$, $get_{dpp2}$, $get_{rancp1}$,\\
&  $get_{rancp2}$, $get_{cncp1}$, $get_{cncp2}$, $get_{ussfp}$, \\
&  $get_{upfp1}$, $get_{upfp2}$$>$ \\ 
\hline
\end{tabular}
\vspace{-0.3cm}
\end{table}

\par Let us consider an NF, for example, CN-DP (UPF)-1, to understand the modelling of the system NF. Various messages (actions) are associated with this NF (CN-DP (UPF)-1) during the session establishment. It has two states, i.e., $Upf_1$ and $Upf_2$. The first state, $Upf_1$, describes the request ($req_{n4est1}$) received from CN controller-1 to establish the N4 session. The second state, $Upf_2$, is for accessing the processor ($get_{upfp1}$) to process the received request and send the response ($rep_{n4est1}$) to CN controller-1 for N4 session establishment. 
\par Each NF requires processing capability to process a request. Therefore, each NF is assigned a corresponding processor, as defined in \cite{R16, R17}. Processors (such as UE processor (UEP), DP-1 processor (DPP), RANC-1 processor (RANCP), USSF processor (USSFP), CNC-1 processor (CNCP) and CN-DP (UPF)-1 processor (UPFP)) are defined using a two-state model for a single processing NF. For instance, the CN-DP (UPF)-1 processor is defined in two states. The first state, $Upfp_1$, to get access to the processor ($get_{upfp1}$), and the second state to perform actions associated with the processor ($rep_{n4est1}$). Similarly, other processors corresponding to their NFs are defined.
\par The system equation describes the overall interaction between the NFs. These interactions are defined as different actions (for example, $S$ = $<$$action_1$, $action_2$$>$) performed between the network functions (say, $NF_1$[N] $_{S}^{\bowtie}$ $NF_2$[N]) to implement the system equation. For example, $Ussf_1$[N] $_{S_2}^{\bowtie}$ $Ranc_1$[N] signifies the interaction between USSF and RANC NFs, where $S_2$ consists of $<$$drb_1$, $notify_1$$>$ actions to interact between these two NFs. Likewise, the interaction between various other NFs is modelled and is shown in Table\,\ref{tab:table1}. In the system equation, \textit{n} is the number of UEs. For the proposed architecture, $N_{nf}$ denotes the number of network functions for a particular category, for example, $N_{dp1}$, $N_{ranc1}$, $N_{ussf}$, $N_{cnc1}$, $N_{upf1}$ denote the number of RAN-DP-1, RAN-controller-1, UE signalling service function, CN controller-1 and CN-DP (UPF)-1 NFs, respectively. Note that each processor can handle a set of concurrent threads, $N_t$ and the number of processors for each network function is denoted as $N_{nfp}$. Thus, $N$ = $N_{nf}$·$N_{nfp}$·$N_t$ (mentioned in the system model equation) represents the total number of threads for an NF of a particular category. Moreover, $N_p$ = $N_{nf}$·$N_{nfp}$ is the total number of processors allocated to a particular NF type. Please note that the table presents only the system model to access one slice, as the modelling remains the same to access another slice. Similarly, modelling is done for existing 5GS procedures in a sliced environment. However, the simulations are performed for both existing 5GS and the proposed architecture considering two slices.

\section{Performance Evaluation}
\label{perf}
This section presents the performance evaluation of the proposed end-to-end slicing network solution. We have created two slices in each of the existing 5GS and in the proposed architecture and then we have evaluated the performance of the session establishment procedure in a sliced environment. We compare the existing 5GS and the proposed architecture based on various performance measures such as the number of sessions established per unit of time slice-wise session establishment rate), Average Response Time (ART), and processor utilization in a sliced environment. The evaluation of these measures also helps in analysing the network's scalability. 
\par Slice-wise session establishment rate measures the frequency of established sessions in the context of the specific action (say, $rep_{se1}$, which represents the response of the request sent from UE for session establishment). ART measures the UE's waiting time for PDU session establishment \cite{R16}. Processor utilization measures the NF's processor capacity utilized during the entire process. 
\par We can see that the proposed architecture with separate controllers, user plane functions and UE signalling service functions can be considered a distributed system similar to the existing 5GS. Hence, we can use the scalability metric of a distributed system to evaluate the scalability of the proposed architecture and compare it with existing 5GS. The scalability metric for a distributed system is based on productivity as defined in \cite{R18}. Therefore, scalability \textit{(Q)} (given in Equation \ref{eq:eq1}) is defined as the ratio between the productivity of a system at two configurations having different scales \textit{$m_1$} and \textit{$m_2$} \cite{R15}. The scaled configurations (\textit{$m_1$} and \textit{$m_2$}) correspond to the different number of NFs used in the network, say (\textit{$m_1$} = (1,1,1,1,1,1,1,1,1) and (\textit{$m_2$} = (3,3,3,3,3,3,3,3,3)). Here, configuration \textit{$m_1$} implies that \textit{($N_{dp1}$, $N_{dp2}$, $N_{ussf}$, $N_{ranc1}$, $N_{ranc2}$, $N_{cnc1}$, $N_{cnc2}$, $N_{upf1}$, $N_{upf2}$)} = \textit{(1,1,1,1,1,1,1,1,1)} for the proposed architecture, which is the basic configuration with single network function assigned for all functions. Similarly, \textit{($N_{dp1}$, $N_{dp2}$, $N_{ussf}$, $N_{ranc1}$, $N_{ranc2}$, $N_{cnc1}$, $N_{cnc2}$, $N_{upf1}$, $N_{upf2}$)} = \textit{(3,3,3,3,3,3,3,3,3)}, which is the configuration for a scaled system. Further, the mathematical expression for scalability is given as \cite{R15}:
    \begin{equation}\label{eq:eq1}
Q(m_1,m_2) = \frac{P(m_2)}{P(m_1)}
\end{equation}  
Where \textit{P(m)} is the productivity of a system at the scale \textit{m} which can be defined as (Equation \ref{eq:eq2}): 
\begin{equation}\label{eq:eq2}
P(m) = \frac{t(m)f(m)}{R(m)}
\end{equation}
Where \textit{t(m)} is the average number of PDU sessions established at scale \textit{m}, \textit{R(m)} is the processor utilization of the system at scale \textit{m}, and \textit{f(m)} (Equation \ref{eq:eq3}) is determined by evaluating the response time performance of the scaled system. We consider the following equation \cite{R18} to evaluate the performance function \textit{f(m)} by using the  average response time \textit{T(m)}, at scale \textit{m}, with the target average response value \textit{T} \cite{R15}: 

\begin{equation}\label{eq:eq3}
f(m) =\frac{1}{1+T(m)/T}
\end{equation}
\begin{figure}[h!]
	\centering
	\includegraphics[width=0.95\columnwidth]{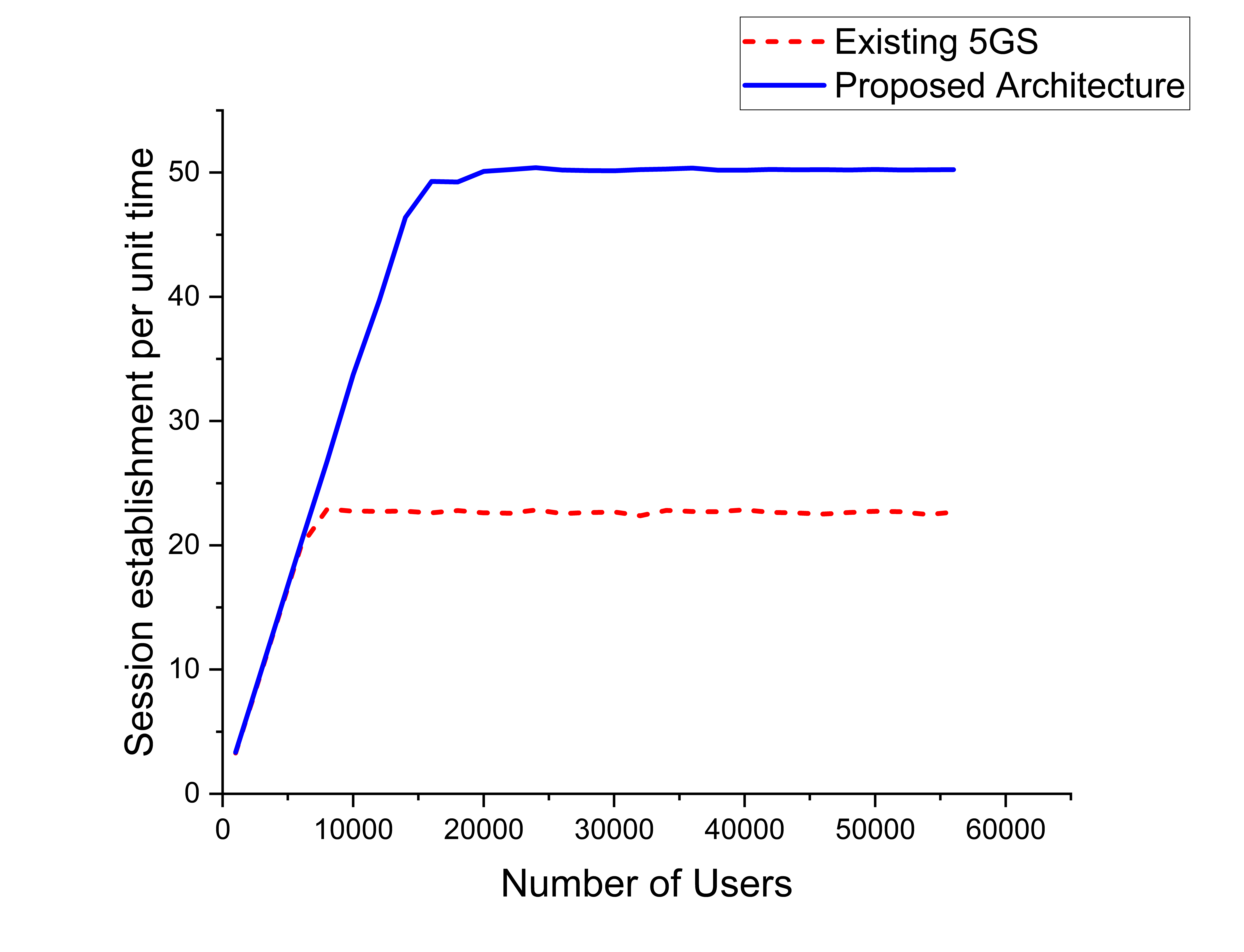}
	\vspace*{-0.3cm}
	\caption{Number of sessions established per unit for UEs  to access network slices for the proposed architecture having the basic configuration of \textit{($N_{dp1}$, $N_{dp2}$, $N_{ussf}$, $N_{ranc1}$, $N_{ranc2}$, $N_{cnc1}$, $N_{cnc2}$, $N_{upf1}$, $N_{upf2}$)} = \textit{(1,1,1,1,1,1,1,1,1)} and for the existing 5GS architecture with the basic configuration of \textit{($N_{du1}$, $N_{du2}$, $N_{cu1}$, $N_{cu2}$, $N_{amf}$, $N_{smf1}$, $N_{smf2}$, $N_{upf1}$, $N_{upf2}$)} = \textit{(1,1,1,1,1,1,1,1,1)}.}
	\label{serv_throu_1}
	\vspace{-0.3cm}
\end{figure}
 \par
\begin{figure}[h!]
	\centering
 \vspace{-0.1cm}
\includegraphics[width=0.95\columnwidth]{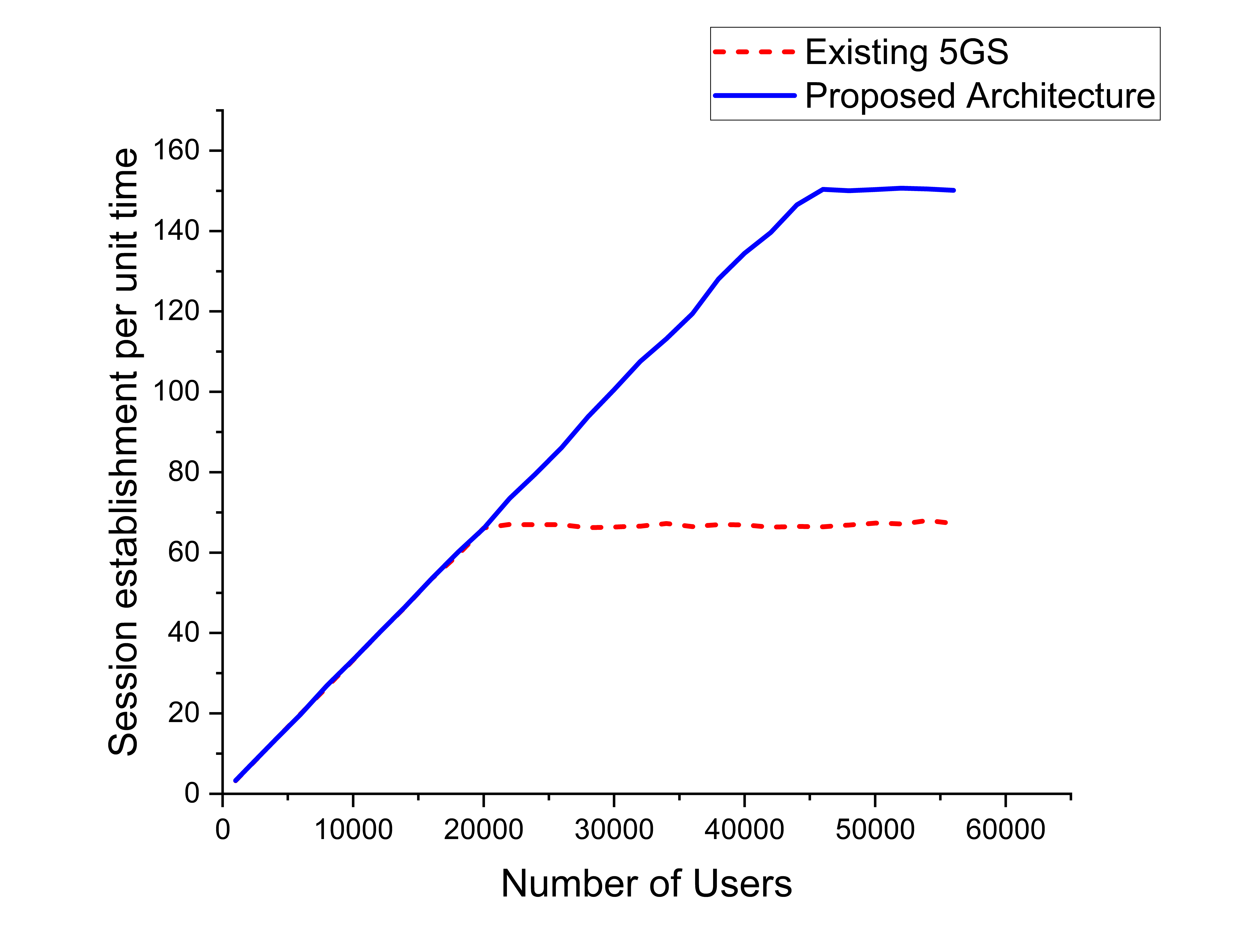}
	\vspace*{-0.4cm}
	\caption{Number of sessions established per unit time to access network slice for the proposed architecture having the scaled configuration of \textit{($N_{dp1}$, $N_{dp2}$, $N_{ussf}$, $N_{ranc1}$, $N_{ranc2}$, $N_{cnc1}$, $N_{cnc2}$, $N_{upf1}$, $N_{upf2}$)} = \textit{(3,3,3,3,3,3,3,3,3)} and for the existing 5GS architecture with the scaled configuration of \textit{($N_{du1}$, $N_{du2}$, $N_{cu1}$, $N_{cu2}$, $N_{amf}$, $N_{smf1}$, $N_{smf2}$, $N_{upf1}$, $N_{upf2}$)} = \textit{(3,3,3,3,3,3,3,3,3)}.}
	\label{serv_throu_3}
	\vspace{-0.2cm}
\end{figure}

 Figures \ref{serv_throu_1} and \ref{serv_throu_3} present the results of the number of sessions established per unit time for UEs to access the network slices in case of the proposed and the existing 5GS architectures for two different configurations (\textit{$m_1$}, \textit{$m_2$}). Even if the same configuration (\textit{$m_1$}) having similar hardware requirements is used for both architectures, the saturation point for existing 5GS is at 8000 users and for the proposed architecture is at 20,000 users for basic configuration, shown in Fig.\,\ref{serv_throu_1}. Similarly, Fig.\,\ref{serv_throu_3} shows that for scaled configuration, existing 5GS saturates at 20,000 users, while the proposed architecture saturates at 46,000 users. Here saturation point is the maximum number of users that can be served by the system. This saturation point corresponds to the saturation of processor utilization (discussed next).

 \begin{figure}[h!]
	\centering
 \vspace{-0.2cm}
\includegraphics[width=\columnwidth]{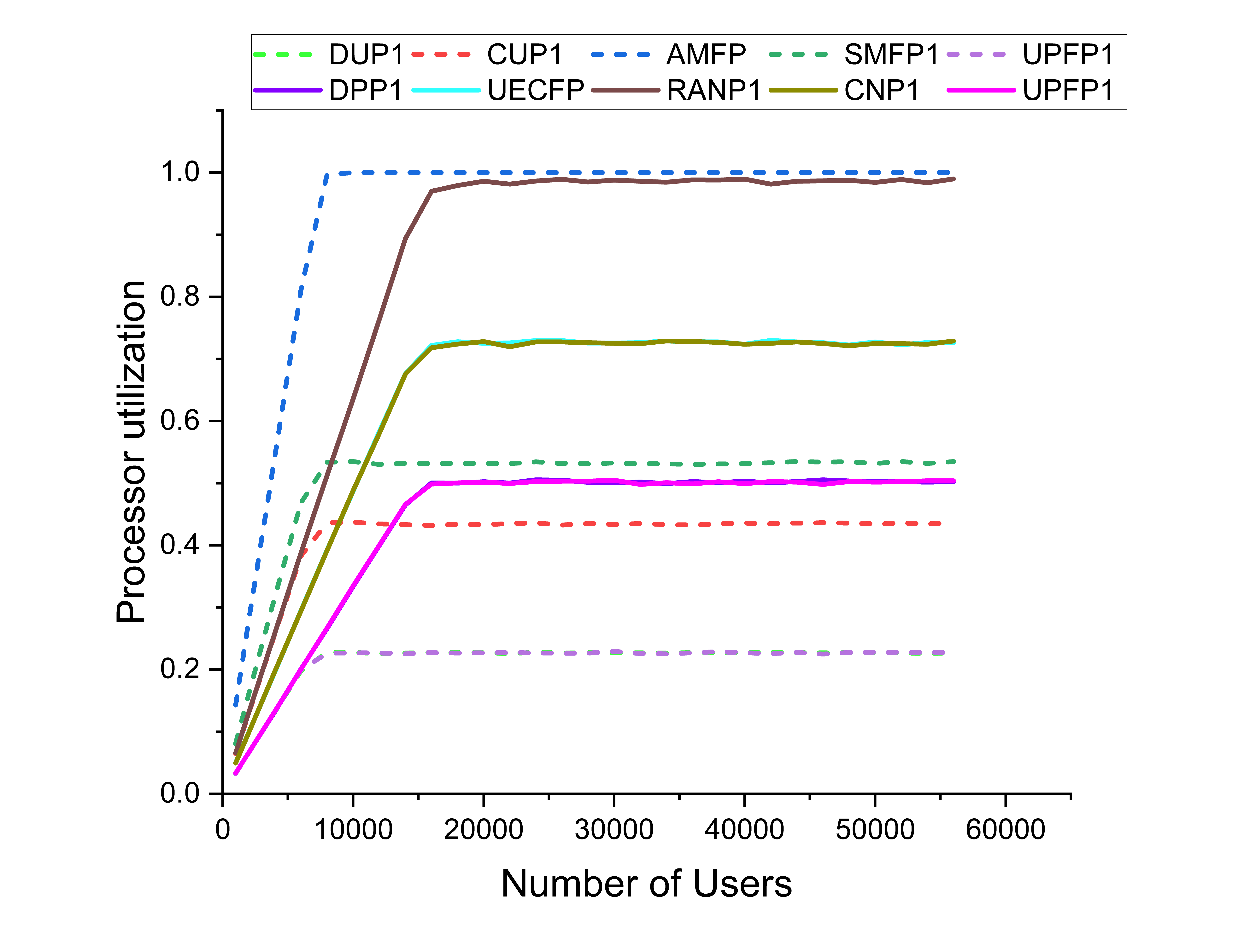}
	\vspace*{-1cm}
	\caption{Processor utilization for the proposed architecture having the basic configuration of \textit{($N_{dp1}$, $N_{dp2}$, $N_{ussf}$, $N_{ranc1}$, $N_{ranc2}$, $N_{cnc1}$, $N_{cnc2}$, $N_{upf1}$, $N_{upf2}$)} = \textit{(1,1,1,1,1,1,1,1,1)} and for the existing 5GS architecture with the basic configuration of \textit{($N_{du1}$, $N_{du2}$, $N_{cu1}$, $N_{cu2}$, $N_{amf}$, $N_{smf1}$, $N_{smf2}$, $N_{upf1}$, $N_{upf2}$)} = \textit{(1,1,1,1,1,1,1,1,1)}.}
	\label{serv_util_1}
	\vspace*{-0.1cm}
\end{figure}
\begin{figure}[h!]
	\centering
 \vspace{-0.3cm}
\includegraphics[width=\columnwidth]{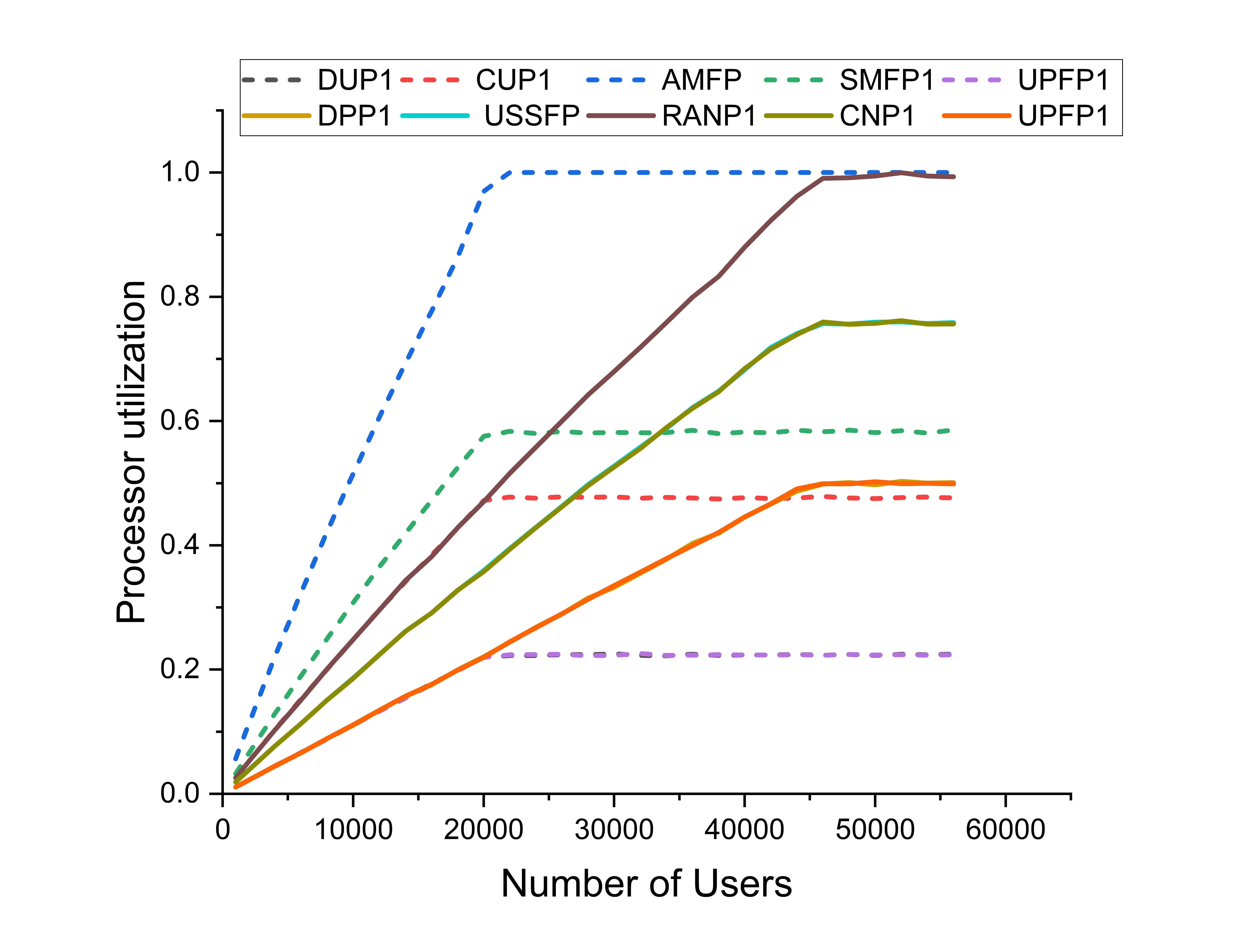}
	\vspace*{-0.8cm}
	\caption{Processor utilization for the proposed architecture having the scaled configuration of \textit{($N_{dp1}$, $N_{dp2}$, $N_{ussf}$, $N_{ranc1}$, $N_{ranc2}$, $N_{cnc1}$, $N_{cnc2}$, $N_{upf1}$, $N_{upf2}$)} = \textit{(3,3,3,3,3,3,3,3,3)} and for the existing 5GS architecture with the scaled configuration of \textit{($N_{du1}$, $N_{cu1}$, $N_{amf}$, $N_{smf1}$, $N_{smf2}$, $N_{upf1}$, $N_{upf2}$)} = \textit{(3,3,3,3,3,3,3,3,3)}.}
	\label{serv_util_3}
	\vspace{0.0cm}
\end{figure}
\begin{figure}[h!]
	\centering
 \vspace{-0.3cm}
\includegraphics[width=\columnwidth]{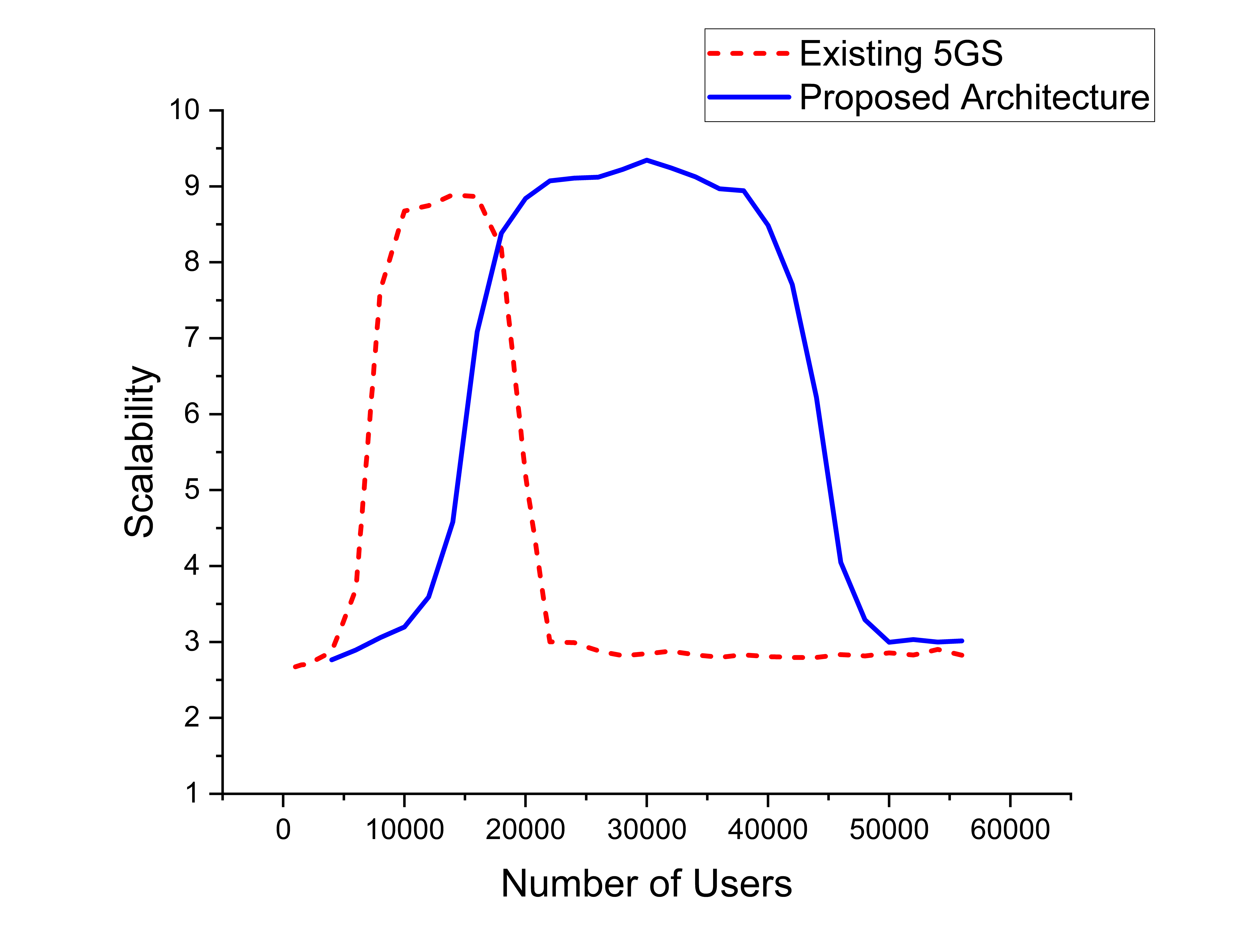}
	\vspace*{-0.8cm}
	\caption{Scalability for the proposed and the existing 5GS architecture.}
	\label{scalability}
	\vspace{-0.5cm}
\end{figure}
\par The processor utilization (for all the NFs of slice-1) of the existing 5GS and the proposed architecture for basic configuration is shown in Fig.\,\ref{serv_util_1}. For instance, the AMFP reaches its maximum utilization explaining the saturation point for the number of session establishments. Although at this point, other NFs are not fully utilized. These results show that the request processing chain fails if an NF becomes a bottleneck for the consecutive chain. Similarly, Fig.\,\ref{serv_util_3} shows the processor utilization (for all the NFs of slice-1) results for the existing 5GS and the proposed architecture by considering scaled configuration. Please note that the solid line in Fig.\,\ref{serv_util_1} and Fig.\,\ref{serv_util_3} represents the processors of the proposed architecture, while the dotted line is for the existing 5GS architecture. It is evident that processors are saturated earlier in the case of existing 5GS as compared to the proposed architecture, as the number of messages in existing 5GS is more compared to the proposed architecture.
 \par Based on results obtained for the slice-wise session establishment rate, ART and processor utilization (from the PEPA-based simulation and modelling), scalability is evaluated (from equation \ref{eq:eq1}) and plotted in Fig.\,\ref{scalability}. We consider the same two configurations \textit{$m_1$} and \textit{$m_2$} as referred to above for the estimation of scalability. The same saturation points can be observed in Fig.\,\ref{scalability}. The existing 5GS saturates at 8,000 users, while the proposed architecture saturates at 20,000 users for basic configuration. Similarly, for scaled configuration, the existing 5GS saturates at 20,000 users, while the proposed architecture saturates at 46,000 users. It indicates that the proposed architecture performs better than the existing 5GS and can serve more concurrent users with the same scaling configuration. Further, it signifies that with the help of the same configuration or resources assigned to both architectures, the proposed one performs better in terms of slice-wise session establishment rate, ART and processor utilization; and as a result, is more scalable. We provide results for the performance of two architectures in the sliced environments using multiple slices. The proposed architecture performs better compared to the existing 5GS architecture in a sliced environment. 

\section{Conclusion} 
\label{conc}
In this paper, we have proposed architectural enhancement for end-to-end slicing by employing slice-specific control plane evolution for beyond 5G networks. In the evolved control plane, the UE signalling service function is responsible for signalling exchange with UEs, which has been decoupled from the existing control plane for efficient slice-specific control function deployment. This kind of slice-specific control plane deployment is not possible in the existing 5GS. For example, gNB-CU-CP typically needs to manage multiple slices simultaneously, whereas, in our proposal, a RAN controller can manage an individual slice in RAN. It also leads to a reduced number of messages and simplified interfaces between the control plane and the data plane. The performance of the existing 5GS and the proposed architecture has been compared based on parameters such as slice-wise session establishment rate, ART, processor utilization and scalability to validate the advantages of the proposed idea. The proposed architecture results in simplified slice-specific control signalling, enhanced modularity of the control plane and improved scalability compared to the existing 5GS. 
\section*{Acknowledgment}
We acknowledge the Ministry of Electronics and Information Technology (MeitY), Govt. of India for supporting the project.





\bibstyle{IEEEtran}
\bibliography{tnsm}

\end{document}